\documentclass[12pt]{article}
\input{epsf}
\begin{document}
%
%
\def\mevc {\ifmmode {\rm MeV}/c \else MeV$/c$\fi}
\def\mevcc {\ifmmode {\rm MeV}/c^2 \else MeV$/c^2$\fi}
\def\gevc {\ifmmode {\rm GeV}/c \else GeV$/c$\fi}
\def\gevcc {\ifmmode {\rm GeV}/c^2 \else GeV$/c^2$\fi}
\def\ra   {\rightarrow}
\newcommand{\Bs} {\ifmmode B_{\mbox{\sl s}}^{0}
                       \else $B_{\mbox{\sl s}}^{0}$\fi}
\newcommand{\Ds} {\ifmmode D_{\mbox{\sl s}}^{+}
                       \else $D_{\mbox{\sl s}}^{+}$\fi}
\newcommand{\dms} {\ifmmode \Delta m_{\mbox{\sl s}} \else 
                           $\Delta m_{\mbox{\sl s}}$\fi}
\newcommand{\xs} {\ifmmode x_{\mbox{\sl s}} \else 
                           $x_{\mbox{\sl s}}$\fi}
\begin{titlepage}
\vspace{3 ex}
%
%
\begin{center}
{
\LARGE \bf \rule{0mm}{7mm}{\boldmath CDF - Run\,II Status and Prospects}\\
}

\vspace{4ex}

{\large
Manfred Paulini \\
}
\vspace{1 ex}

{\em
Carnegie Mellon University, Pittsburgh, Pennsylvania 15213, U.S.A. \\
}
\vspace{2 ex}
%
%
{\large
For the CDF Collaboration
}
\end{center}

\vspace{2 ex}
%
%
\begin{abstract}
After a five year upgrade period, the CDF detector located at the
Fermilab Tevatron Collider is back in operation
taking high quality data with all subsystems functional.
We report on the status of the CDF experiment in Run\,II and
discuss the start-up of the Tevatron accelerator.
First physics results from CDF are presented. We also
discuss the prospects for $B$~physics in Run\,II, in particular 
the measurements of \Bs~flavour oscillations and $CP$~violation in
$B$~decays.  
\end{abstract}

\end{titlepage}

%
\setlength{\oddsidemargin}{0 cm}
\setlength{\evensidemargin}{0 cm}
\setlength{\topmargin}{0.5 cm}
\setlength{\textheight}{22 cm}
\setlength{\textwidth}{16 cm}
\setcounter{totalnumber}{20}
\topmargin -1.0cm

\clearpage\mbox{}\clearpage

\pagestyle{plain}
\setcounter{page}{1}
%
\subsection*{Introduction}

The CDF experiment can look back to a successful $B$~physics program during
the 1992-1996 Run\,I data taking period. The highlights of Run\,I
$B$~physics results from 
CDF include the measurements of the lifetimes of all
$B$~hadrons~\cite{myrevart}, several measurements of the $B^0$~flavour
oscillation frequency $\Delta m_d$~\cite{myrevart}, the discovery of the
$B_c$~meson~\cite{bcdisc} and first evidence that the $CP$~violation
parameter $\sin 2\beta$ is different from zero~\cite{sin2b_cdf}.
Since 1996, the CDF~detector has undergone a major upgrade~\cite{cdfup}
to allow operation at high luminosities and bunch spacings of up to 132~ns,
as originally planned for Run\,II of the Tevatron. The upgraded CDF
detector contains several completely new components that took years for
successful completion. 

Run\,II officially started in March 2001. Since then much work has gone
into commissioning the CDF detector. With the beginning of 2002, the CDF
detector is in stable running conditions operating with reliable physics
triggers. 
We report about the first physics results obtained with the
upgraded CDF~detector from the analysis of the first Run\,II data collected
until the summer of 2002. 

\subsection*{Start-up of Run\,II}

\subsubsection*{The Upgraded Tevatron Accelerator}

The Fermilab accelerator complex has also undergone a major upgrade since
the end of Run\,I. The centre-of-mass energy has been increased to
1.96~TeV. But most importantly, the Main Injector, a new 150~GeV
proton storage ring, has replaced the Main Ring as injector of protons and
anti-protons into the Tevatron. The Main Injector also 
provides higher proton intensity onto the anti-proton production target,
allowing for more than an order of magnitude higher luminosities.
An additional new storage ring, the Recycler, housed in the same tunnel as
the Main Injector, will allow to reuse anti-protons at the end of each store. 
The design luminosity during the first phase of Run\,II (Run\,IIa) is 
5-8$\cdot 10^{31}$~cm$^{-2}$s$^{-1}$ for a final integrated luminosity of
$\sim 2$~fb$^{-1}$ by the end of Run\,IIa. The present bunch crossing time
is 396~ns with a $36\times36$ $p\bar p$ bunch operation.

\begin{figure}[tbp]
\centerline{
\epsfysize=6.5cm
\epsffile{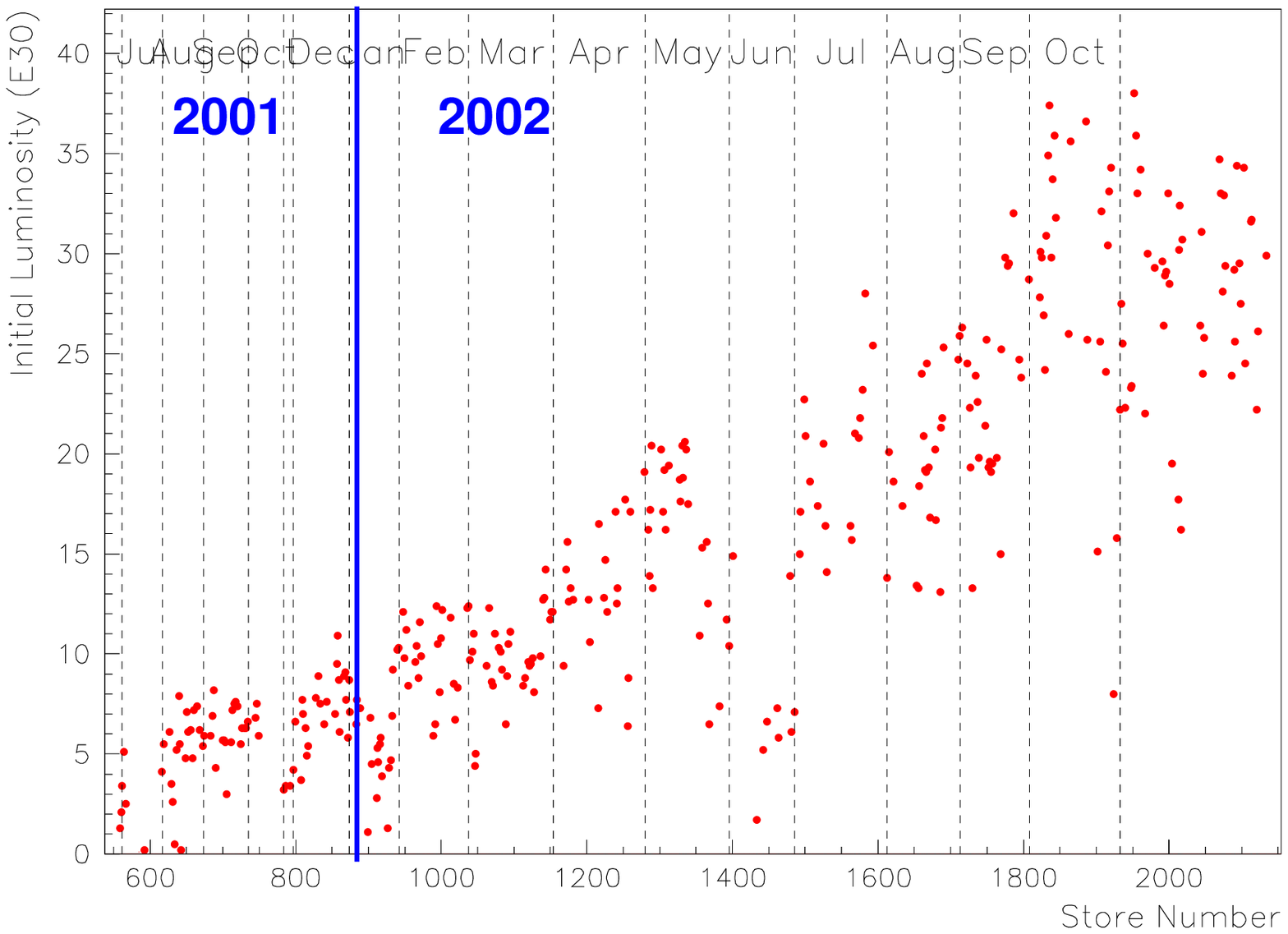}
\epsfysize=6.7cm
\epsffile{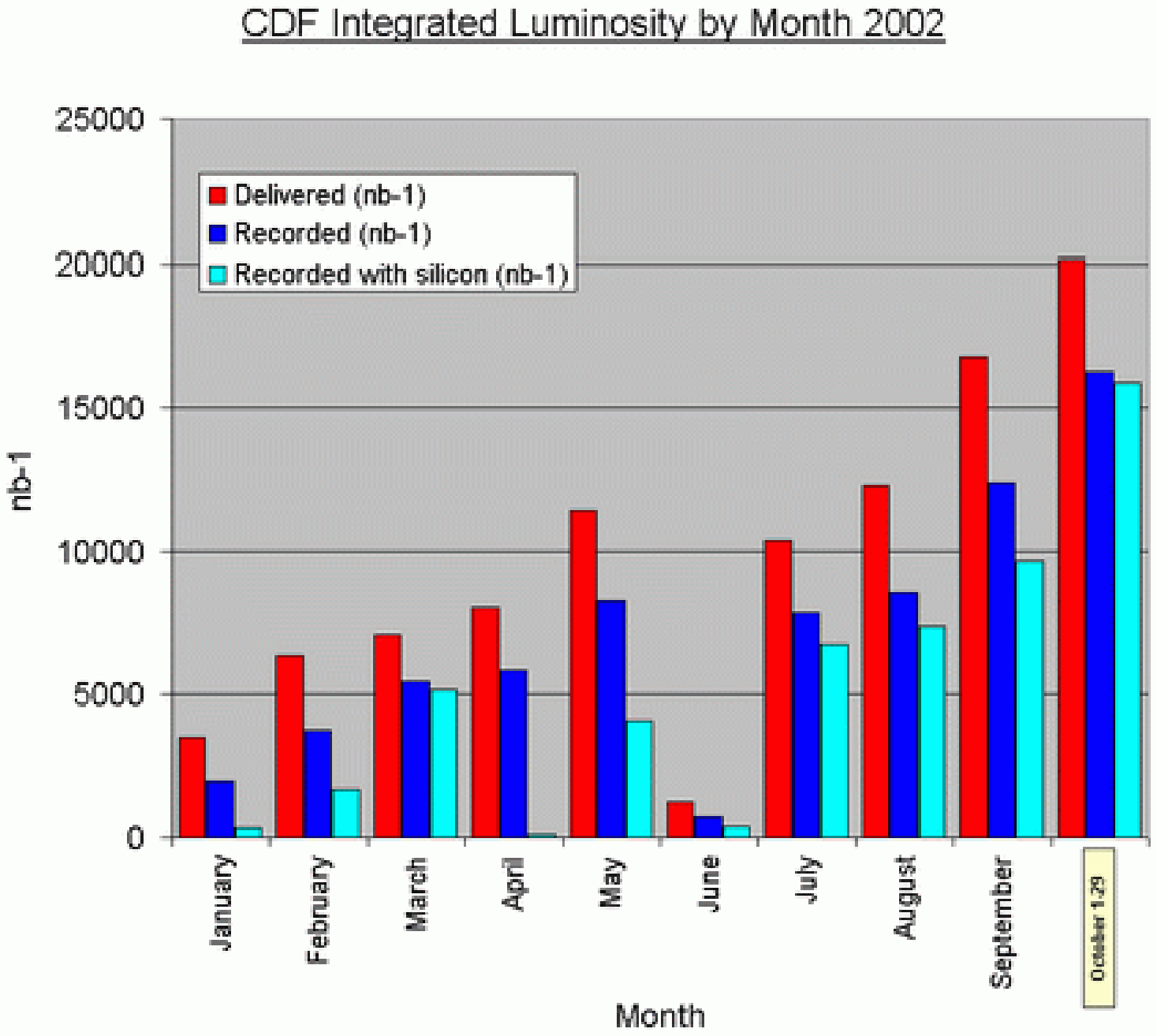}
\put(-425,130){\large\bf (a)}
\put(-230,150){\large\bf (b)}
}
\caption{
Distribution of (a) the initial luminosity delivered by the Tevatron and
(b) the integrated luminosity per month delivered by the Tevatron and
recorded by CDF as well as the fraction of data with silicon information.
}
\label{tev_lumi}
\end{figure}

Figure~\ref{tev_lumi}(a) shows the development of the initial Tevatron
luminosity from late 2001 to late 2002. So far, the peak luminosity reached
by the Tevatron is 3.7$\cdot 10^{31}$~cm$^{-2}$s$^{-1}$ still below
expectations. 
Figure~\ref{tev_lumi}(b) is a diagram of the integrated luminosities
delivered and recorded by CDF. This graph also includes the amount of data
recorded with information from CDF's silicon vertex detector. The
total integrated luminosity recorded by CDF in 2002 is about 100~pb$^{-1}$
(equivalent to the Run\,I data set). 

\subsubsection*{CDF Detector Performance in Run\,II}

The CDF detector improvements for Run\,II~\cite{cdfup} were motivated by
the shorter 
accelerator bunch spacing of up to 132~ns and the increase in luminosity by
an order of magnitude. All front-end and trigger electronics has been
significantly redesigned and replaced. A DAQ upgrade allows the operation
of a pipelined trigger system. CDF's tracking 
system was completely upgraded. It consists of  
a new Central Outer Tracker (COT) with 30,200 sense wires
arranged in 96 layers combined into four axial and four stereo
superlayers. It also provides d$E$/d$x$ information for particle
identification. 
The Run\,II silicon vertex
detector consists of seven double sided layers and one single sided layer
mounted on the beampipe covering a total radial area from 1.5-28~cm. The
silicon vertex detector covers the full Tevatron luminous 
region which has a RMS spread of about 30~cm along the beamline and allows
for standalone silicon tracking up to a pseudo-rapidity $|\eta|$ of 2. The
forward calorimeters have been replaced by a new scintillator tile based
plug calorimeter which gives good electron identification up to 
$|\eta|=2$.
The upgrades to the muon system almost double the central
muon coverage and extent it up to $|\eta|\sim1.5$.

\begin{figure}[tbp]
\centerline{
\epsfysize=7.2cm
\epsffile{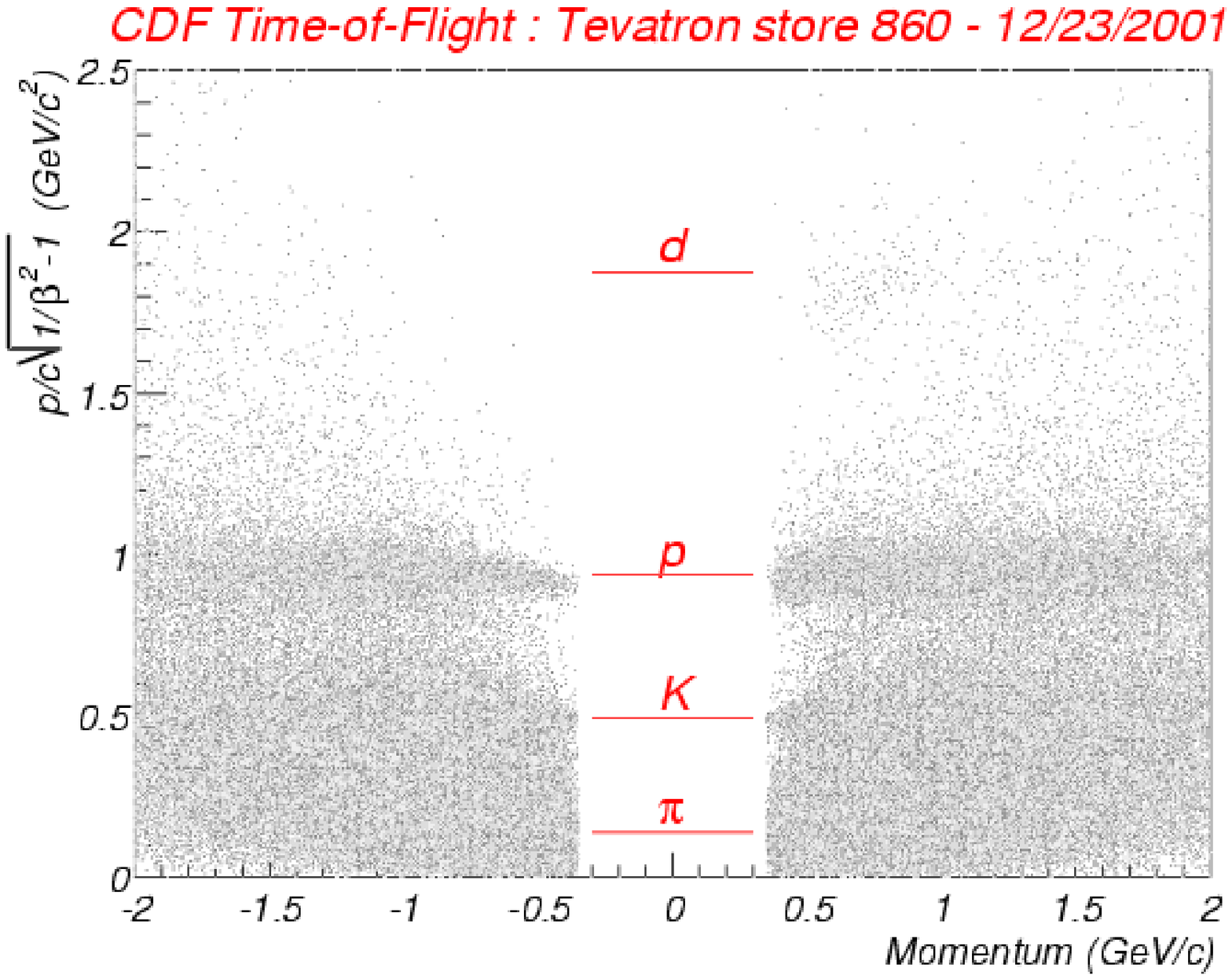}
\epsfysize=7.4cm
\epsffile{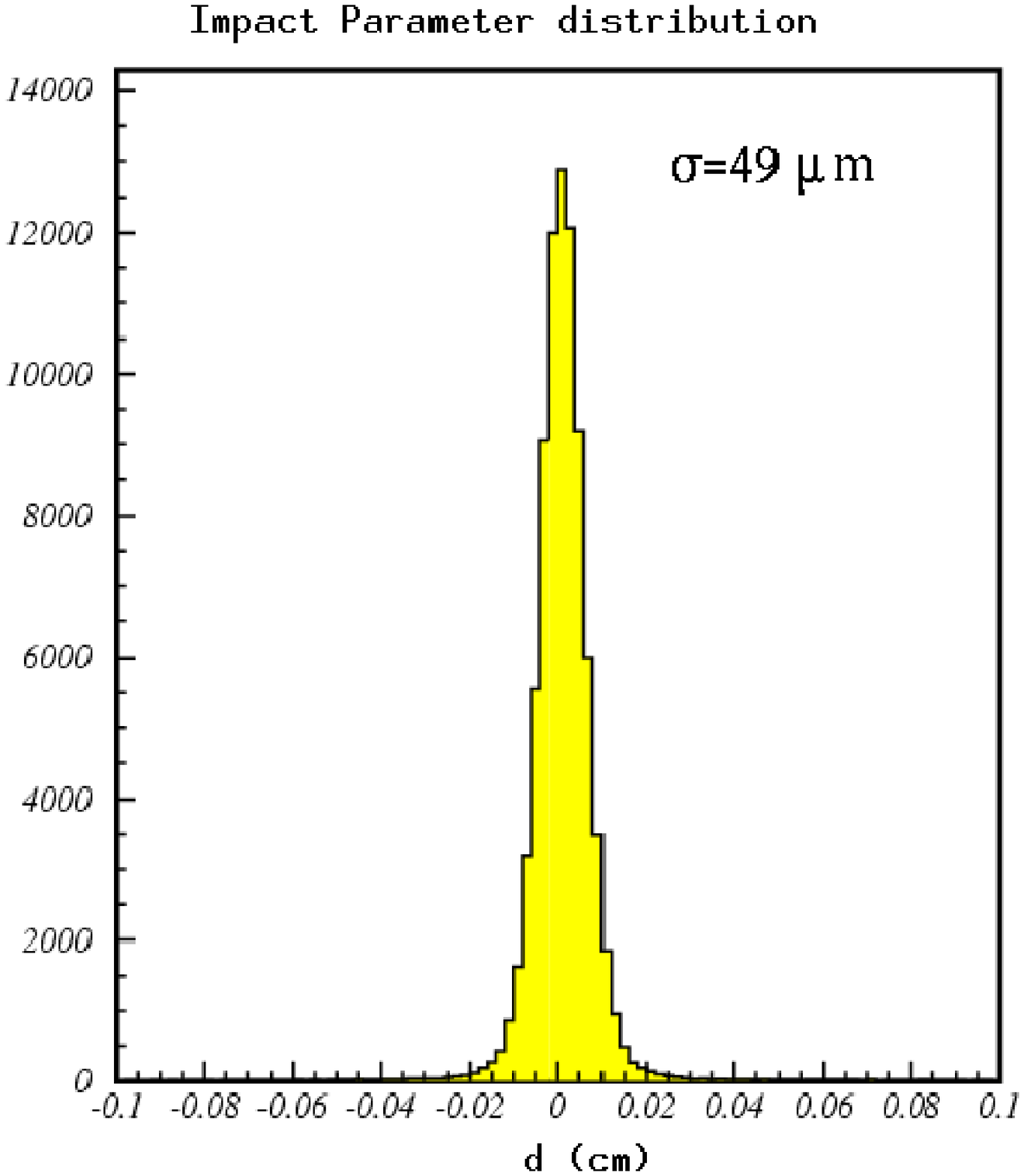}
\put(-395,160){\large\bf (a)}
\put(-140,170){\large\bf (b)}
}
\caption{
Distribution of (a) the Time-of-Flight reconstructed mass versus momentum
for positive and negative tracks and (b) SVT track impact parameter
distribution.  
}
\label{tof_svt}
\end{figure}

The most important improvements for $B$~physics in Run\,II are a 
Silicon Vertex Trigger (SVT)
and a Time-of-Flight (ToF) system with a resolution of about
100~ps. The later employs 216 three-meter-long
scintillator bars located between the outer radius of the COT
and the superconducting solenoid.
The CDF\,II Time-of-Flight detector and its performance
has been presented by
S.~Giagu~\cite{giagu_tof} at this conference. As an example of the
preliminary performance of  
the new Time-of-Flight system, Figure~\ref{tof_svt}(a) shows the
distribution of reconstructed mass versus momentum for positive and
negative tracks showing clear separation of $\pi$, $K$ and $p$. 
The Time-of-Flight system will be most beneficiary for the identification
of kaons with a 2\,$\sigma$-separation between $\pi$ and $K$ for
$p<1.6$~\gevc. This will enable CDF to make use of opposite side kaon
tagging and allows to identify same side fragmentation
kaons accompanying \Bs~mesons~\cite{giagu_tof}.

In Run\,I, all $B$~physics triggers at CDF were based on leptons including
single and dilepton triggers. The newly implemented Silicon Vertex Trigger
gives CDF access to purely hadronic $B$~decays and makes CDF's $B$~physics
program fully competitive with the one at the
$e^+e^-$~$B$~factories. The hadronic track trigger
is the first of its kind operating successfully at a hadron collider. It
works as follows: With a
fast track trigger at Level~1, CDF finds track pairs in the COT
with $p_T>1.5$~\gevc. At Level~2, these tracks are
linked into the silicon vertex detector and cuts on the track impact
parameter (e.g.~$d > 100$ $\mu$m) are applied. 
The original motivation for CDF's hadronic track trigger was to select
the two tracks from the rare decay $B^0 \ra \pi\pi$. 
The Silicon Vertex Trigger was fully operational at the time of this
conference. A detailed discussion of CDF's Silicon Vertex Trigger and its
initial performance has been presented by D.~Lucchesi~\cite{lucchesi_svt}
at this conference. As an example of the performance of this trigger, we
show the SVT track impact parameter distribution 
which behaves as expected. Figure~\ref{tof_svt}(b)
indicates a resolution of about 50~$\mu$m including a 33~$\mu$m
contribution from the transverse beam spreading.

\subsection*{First Run\,II Physics Results from CDF}

\begin{figure}[tbp]
\centerline{
\epsfysize=6.5cm
\epsffile{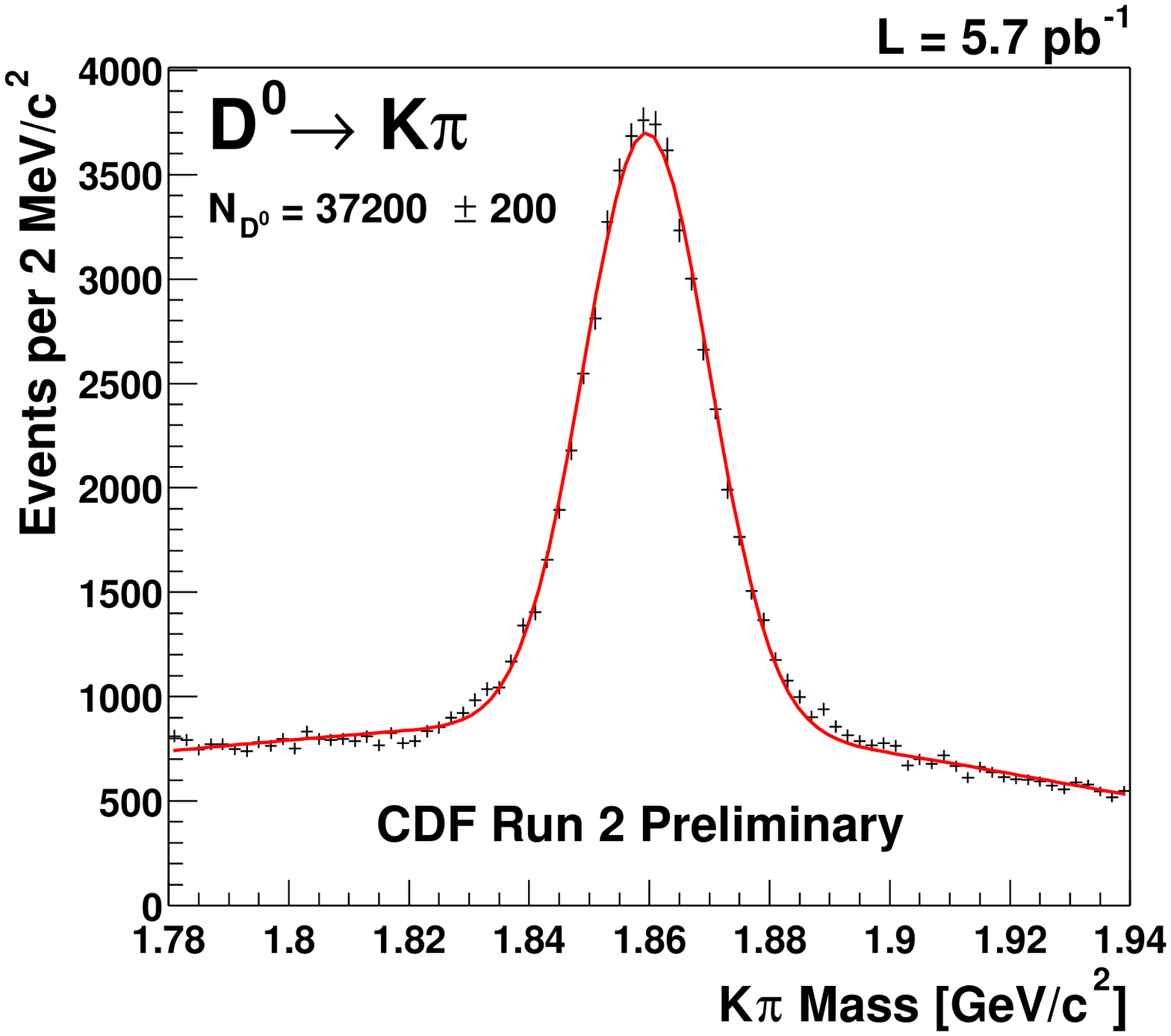}
\epsfysize=6.5cm
\epsffile{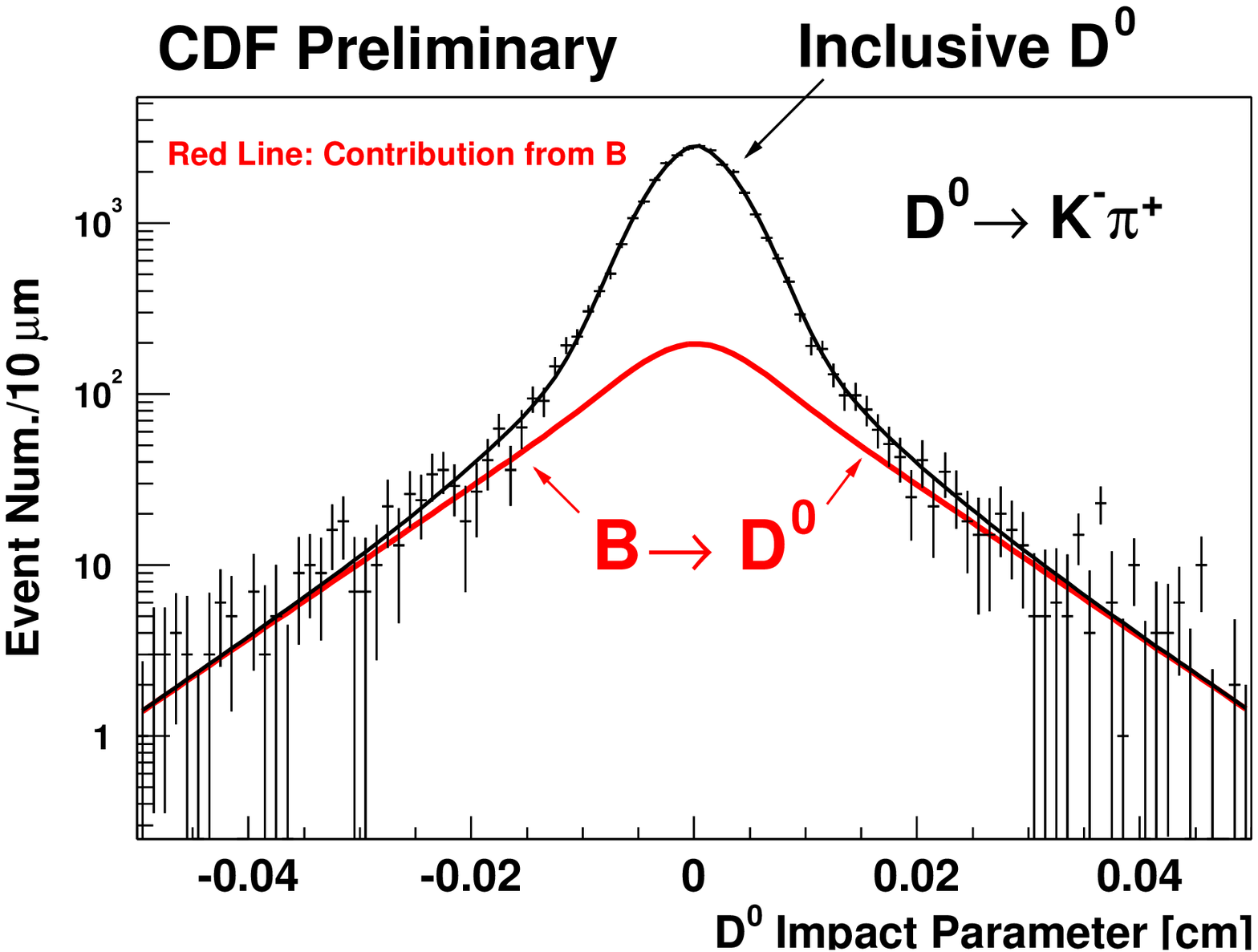}
\put(-415,120){\large\bf (a)}
\put(-200,130){\large\bf (b)}
}
\caption{
(a) $K\pi$ invariant mass from data collected with the SVT
hadronic track trigger. (b) $D^0$ impact parameter distribution indicating
the contribution from $B\ra D^0$ and direct charm.
}
\label{phys1}
\end{figure}

At the time of this conference in June 2002, the performance of the CDF
detector was already close to what has been achieved in the Run\,I data
taking period and the new devices such as the hadronic track trigger were
fully operational. As an example of the
capabilities of the SVT two-track trigger, 
Figure~\ref{phys1}(a) shows
a $K\pi$~mass distribution displaying a
$D^0$ signal of $37,200\pm200$~events obtained in a small data sample of
only 5.7~pb$^{-1}$. The hadronic track trigger clearly collects
large data samples from long-lived charm and beauty decays. 

Where does this charm come from? Is it subsequent
charm from $B$~decays or direct charm?  Figure~\ref{phys1}(b) shows the
impact parameter $d_0$ of the reconstructed $K\pi$-pair. In the case
of direct charm, the $D^0$ points back to the primary vertex resulting in   
$d_0 \sim 0$ while in the case of the $D^0$ originating from a 
$B\ra D^0$~decay,
the $D^0$ does not necessarily extrapolate to the primary interaction
vertex resulting in $d_0 \neq 0$. As can be seen from
Fig.~\ref{phys1}(b), CDF collects large amounts of direct charm with the
hadronic track trigger - a new physics opportunity which did not exist
in Run\,I.

\begin{figure}[tbp]
\centerline{
\epsfysize=6.8cm
\epsffile{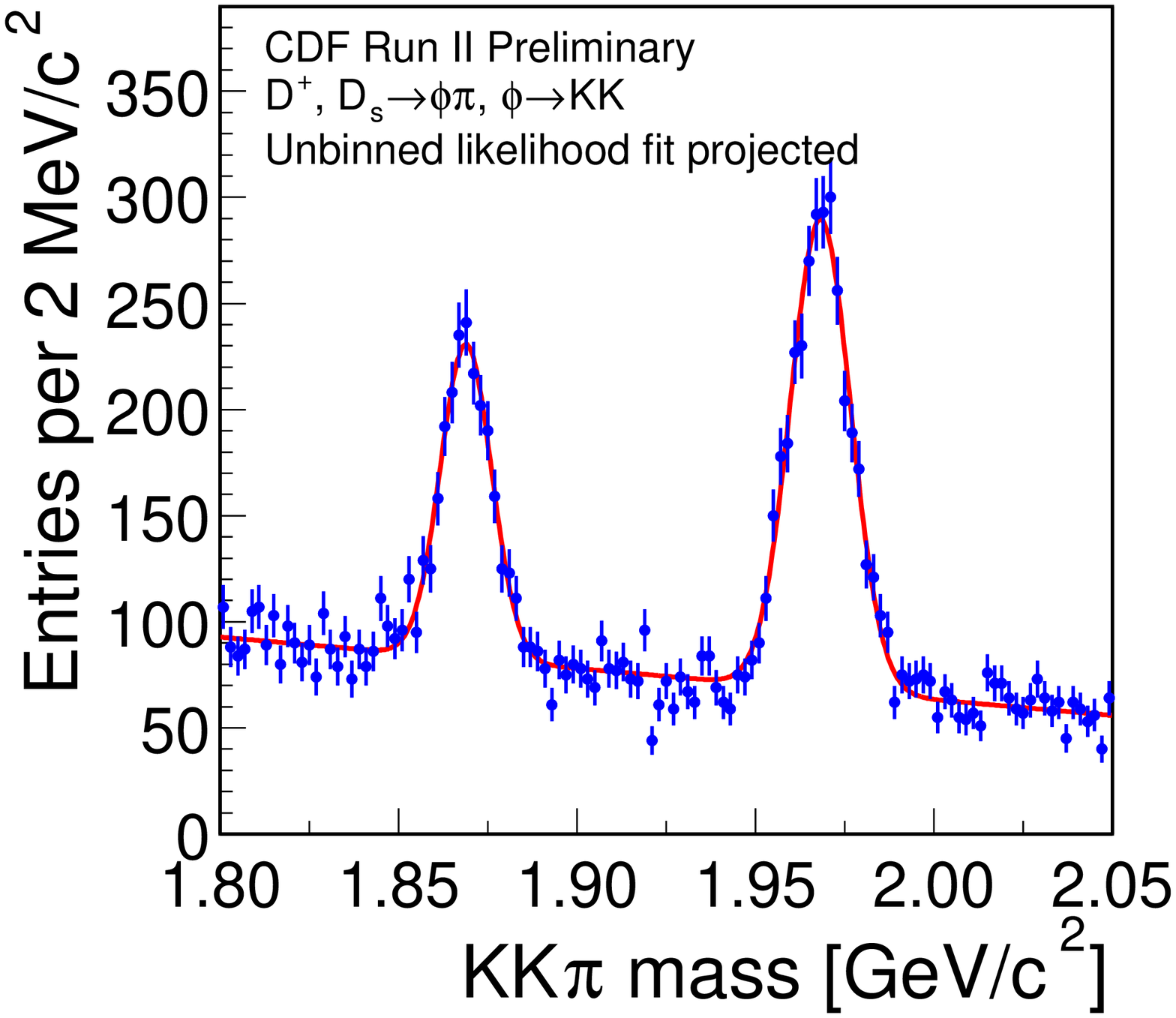}
\epsfysize=6.7cm
\epsffile{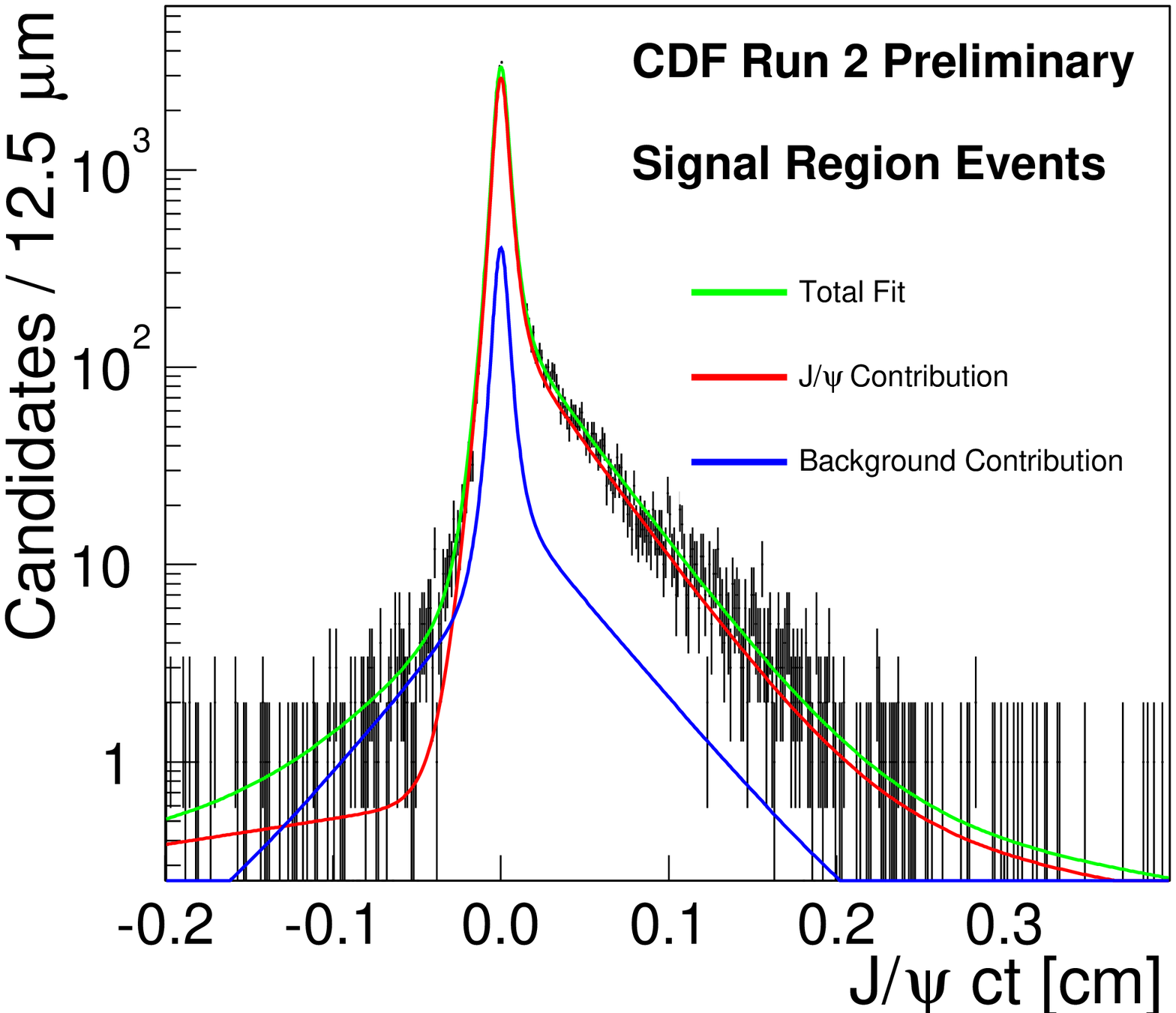}
\put(-395,140){\large\bf (a)}
\put(-175,160){\large\bf (b)}
}
\caption{
(a) $KK\pi$ invariant mass for $\Ds/D^+\ra\phi\pi^+$ candidate events.
(b) Decay length distribution of inclusive $J/\psi$~events with the result
of the lifetime fit superimposed.
}
\label{phys2}
\end{figure}

The physics results presented by CDF in the summer of 2002 were based on
a data sample of 10-20~pb$^{-1}$ and intended to demonstrate that the CDF
detector is
working and calibrated. They included, for example, first measurements of
$B$~meson masses such as 
$m(B^+) = (5280.6\pm1.7\pm1.1)$~\mevcc,  
$m(B^0) = (5279.8\pm1.9\pm1.4)$~\mevcc\ and   
$m(\Bs) = (5360.3\pm3.8\pm^{2.1}_{2.9})$~\mevcc, or  
the measurement of the mass difference 
$m(\Ds) - m(D^+) = (99.41\pm0.38\pm0.21)$~\mevcc\ with comparable precision
to the current world average $\Ds/D^+$ mass
difference~\cite{pdg_2002}. For this measurement, the common decay mode 
$\Ds/D^+\ra\phi\pi^+$ was used.
Figure~\ref{phys2}(a) show the  
$KK\pi$ invariant mass for the $\Ds/D^+$ candidate events.
Another example of CDF\,II physics results is the measurement of the
inclusive $B$~lifetime using $J/\psi\ra\mu\mu$ events demonstrating the 
good understanding of CDF's new silicon vertex detector. The obtained
result $c\tau_b = (458\pm10\pm11)\ \mu$m is in good agreement with other
measurements~\cite{myrevart,pdg_2002}.
Figure~\ref{phys2}(b) shows the  
decay length distribution from inclusive $J/\psi$~events with the result
of the lifetime fit superimposed.

\begin{boldmath}
\subsection*{Prospects for $B$~Physics in Run\,II}
\end{boldmath}

The highlights of CDF's expected $B$~physics program in Run\,IIa include the
discovery of \Bs~flavour oscillations, measurements of $CP$~violation in
$B$~decays as well as searches for rare $B$~decays and measurements of
$B$~production and spectroscopy such as $B_c$~physics. Originally a
luminosity of 2~fb$^{-1}$ was anticipated to be available in the first two
years of Run\,II. Some of the prospects presented below are based on such a
data sample.

\begin{boldmath}
\subsubsection*{\Bs~Flavour Oscillations}
\end{boldmath}

$B^0 \bar B^0$ and $\Bs\bar{\Bs}$ flavour
oscillations measure the Cabibbo-Kobayashi-Maskawa matrix elements
$|V_{td}|/|V_{ts}|$. 
The measurement of the \Bs~oscillation frequency \dms\ is clearly one of
the goals for CDF in Run\,IIa. 
Some of the detector
upgrades play an important role in CDF's prospects for measuring
\Bs~mixing.  
The inner layer of silicon mounted on the beampipe
improves the time resolution for measuring the \Bs~decay length 
to 0.045~ps. This will be important if 
\dms\ is unexpectedly large. The Time-of-Flight
system will enhance the effectiveness of $B$~flavour tagging, especially
through same side tagging with kaons and opposite side kaon tagging, to a
total $\varepsilon {\cal D}^2 \sim 11.3\%$~\cite{giagu_tof}. 
CDF expects a signal of about 75,000 fully reconstructed 
$\Bs \ra \Ds \pi^-,\ \Ds \pi^- \pi^+\pi^-$
events 
in 2~fb$^{-1}$. 

\begin{figure}[tbp]
\centerline{
\epsfxsize=8.0cm
\epsffile{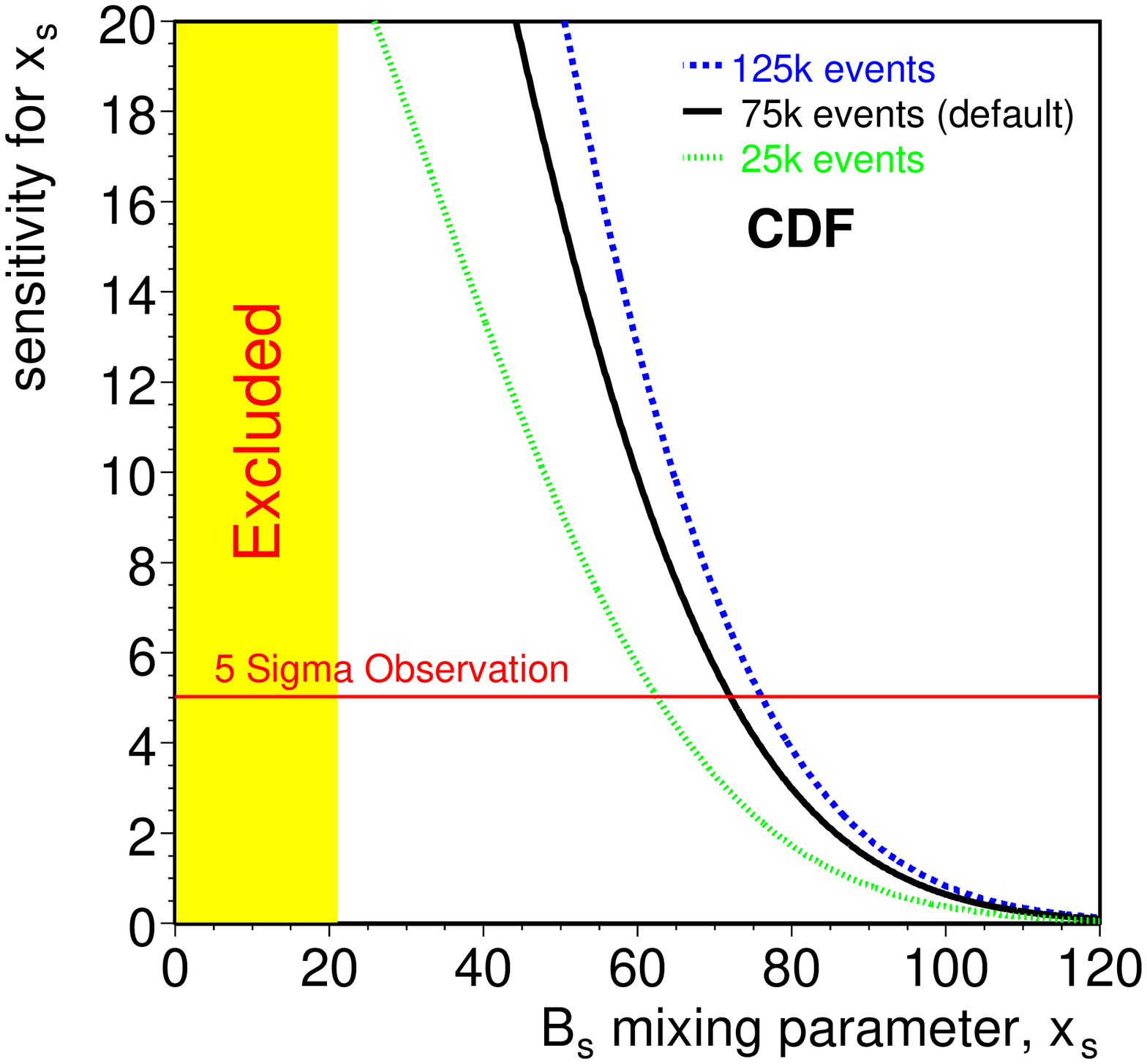}
\epsfxsize=8.0cm
\epsffile{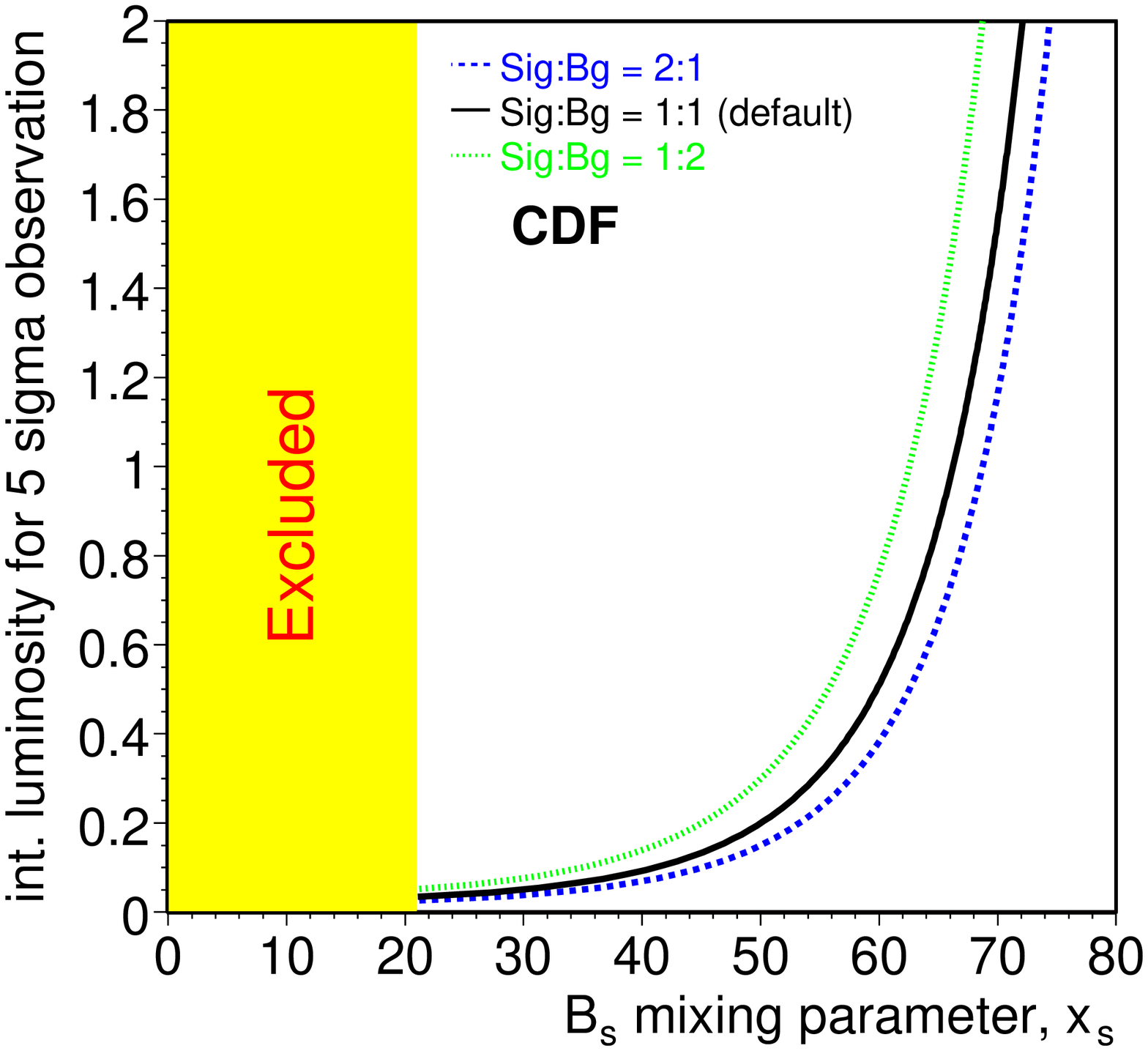}
\put(-280,140){\large\bf (a)}
\put(-130,140){\large\bf (b)}
}
\caption{
(a) Expected sensitivity for a
5\,$\sigma$-observation of \Bs~mixing as a function of the mixing parameter
$\xs$ for various event yields. (b)
Integrated luminosity needed for a 5\,$\sigma$-observation 
versus $\xs$ for different signal-to-background ratios assuming a default
sample size of 75k fully reconstructed \Bs~decays.
}
\label{bsprosp}
\end{figure}

Figure~\ref{bsprosp}(a) shows the expected sensitivity for a
5\,$\sigma$-observation of \Bs~mixing as a function of the mixing parameter
$\xs = \dms/\Gamma$ for various event yields. In Fig.~\ref{bsprosp}(b), the
integrated luminosity needed for a 5\,$\sigma$-observation is plotted
versus $\xs$ for different signal-to-background ratios assuming a sample
size of
75k fully reconstructed \Bs~decays. If the \Bs~mixing frequency is around
the current Standard Model (SM) expectation of $\dms \sim
20$~ps$^{-1}$~\cite{pdg_2002}, Figure~\ref{bsprosp}(b) indicates that CDF
would only need a few hundred pb$^{-1}$ to discover \Bs~flavour
oscillations. This assumes all detector components and triggers work as
expected. There appear to be indications that the projected event yield
might be overestimated. Given this and the 
small amount of data delivered by the Tevatron at the time of this
conference (and even at the end of 2002), it will take some time until CDF
can present first results on \Bs~mixing.
A measurement of \dms\ will be the next crucial test of the SM
probing whether the obtained result will fit to the current constraints on
the CKM 
triangle which are all in beautiful agreement~\cite{pdg_2002}. 
It is noteworthy to mention that physics with \Bs~mesons
is unique to the Tevatron until the start of the LHC in 2007. 

\newpage
\begin{boldmath}
\subsubsection*{$CP$ Violation}
\end{boldmath}

Several years ago, 
the most important decay modes for the study of $CP$~violation in the
$B$~system were
believed to be $B^0\ra J/\psi K^0_S$ and $B^0\ra\pi^+\pi^-$.
The time dependence of $CP$~violation in the former mode measures
$\sin2\beta$, while
the decay $B^0\to\pi^+\pi^-$ usually appears in the literature as 
a tool to determine $\alpha=180^\circ-\beta-\gamma$. 
However, it has been  shown that the
so-called penguin pollution  in $B^0\ra\pi^+\pi^-$
is sufficiently large to make the
extraction of fundamental physics parameters from the measured
$CP$~asymmetry rather difficult. 
An evaluation of measuring $CP$~violation in $B^0\ra\pi^+\pi^-$
does therefore 
require a strategy to distinguish penguin contributions  from 
tree diagrams.
A large number of
strategies to disentangle both contributions is discussed in
the literature~\cite{revs,alph}.
However, they generally require either very large data sets
or involve hard to quantify theoretical uncertainties.

CDF evaluated a strategy of measuring the 
CKM angle~$\gamma$ as suggested by Fleischer in Ref.~\cite{flekk}.
This method is
particularly well matched to the capabilities of the Tevatron as 
it relates $CP$~violating 
observables in 
$\Bs \ra K^+ K^-$ and $B^0\ra\pi^+\pi^-$.
Both decays are related to each other by interchanging all
down and strange quarks, i.e.\ through the so-called ``U-spin''
subgroup of the SU(3) flavour symmetry of strong interactions. 
The strategy proposed in Ref.~\cite{flekk} uses this
symmetry to relate the ratio of hadronic matrix elements
for penguins and trees, and thus uses $\Bs \ra K^+ K^-$ to correct for the
penguin pollution in $B^0\ra\pi^+\pi^-$.

With the hadronic track trigger, CDF expects to reconstruct at least 
5000 $B^0\to\pi^+\pi^-$ and 20,000 $B^0\to K^\pm\pi^\mp$ events in
2~fb$^{-1}$  assuming a branching ratio of
${\cal B}(B^0\to \pi^+\pi^-)=5\times 10^{-6}$.
The question whether
CDF will be able to extract these signals from potentially enormous
backgrounds, has been studied.
With respect to combinatorial background, a signal-to-background ratio not
worse than $S/B \sim 0.4$ can be expected.
Regarding physics backgrounds from $B \ra K \pi$ and $\Bs \ra K K$ decays,
a $B^0\ra\pi^+\pi^-$ signal can be extracted by exploiting
the invariant $\pi\pi$ mass 
distribution as well as the d$E$/d$x$ information provided by CDF's
Central Outer Tracker. From this, CDF expects the $B\ra \pi\pi$, $K\pi$,
$KK$ and $\pi K$ yields to be measured with an
uncertainty of only a few percent.

Measurements on the tagged samples determine the time dependent
$CP$~asymmetry for $B^0\to \pi^+\pi^-$ and $\Bs\to K^+K^-$ which is given by:
$
{\cal A}_{CP} = {\cal A}^{dir}_{CP} \cos \Delta m t + 
{\cal A}_{CP}^{mix} \sin \Delta m t  
$.
With the strategy suggested in Ref.~\cite{flekk}, studies at CDF 
indicate that a measurement of the CKM angle
$\gamma$ to better than $10^{\circ}$ could be feasible with
2~fb$^{-1}$ of data. 
The utility of these modes depends on how well the uncertainty
from flavour $SU(3)$ breaking can be controlled.
Data for these and other processes should tell us the range of such
effects. The resulting Standard Model constraints could be quite
stringent. 
CDF estimates of possible SU(3) breaking effects show that 20\% SU(3)
breaking leads to a systematic error of less than half the statistical
precision given above. This encouraging result might allow CDF to make a
significant contribution to our understanding of the CKM unitarity triangle
within the first 2~fb$^{-1}$ of Tevatron data in Run\,II.

\vskip 0.4cm
\begin{boldmath}
\subsubsection*{Rare $B$ Decays}
\end{boldmath}

\begin{figure}[tbp]
\centerline{
\epsfxsize=8.0cm
\epsffile{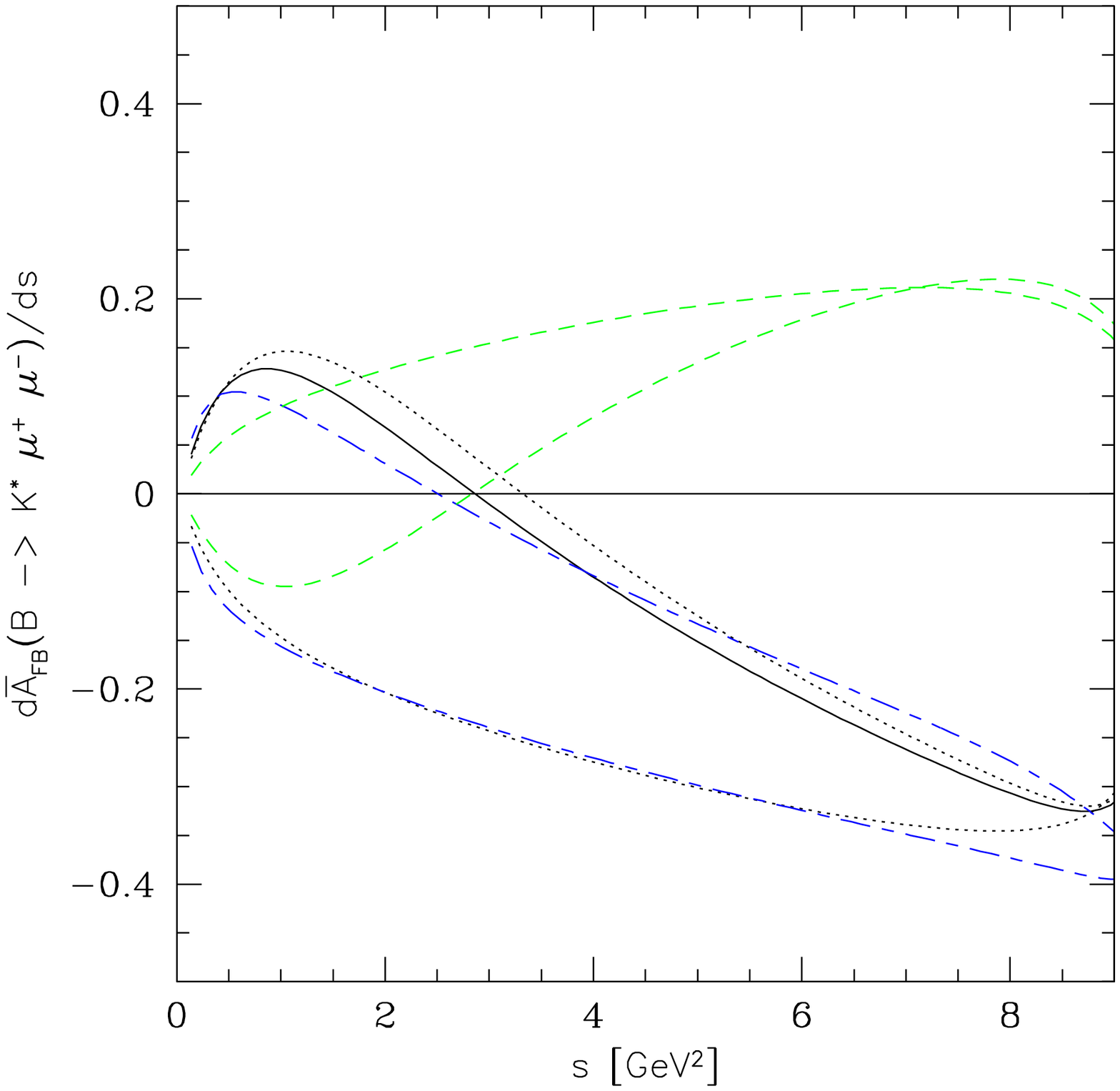}
\epsfxsize=8.0cm
\epsffile{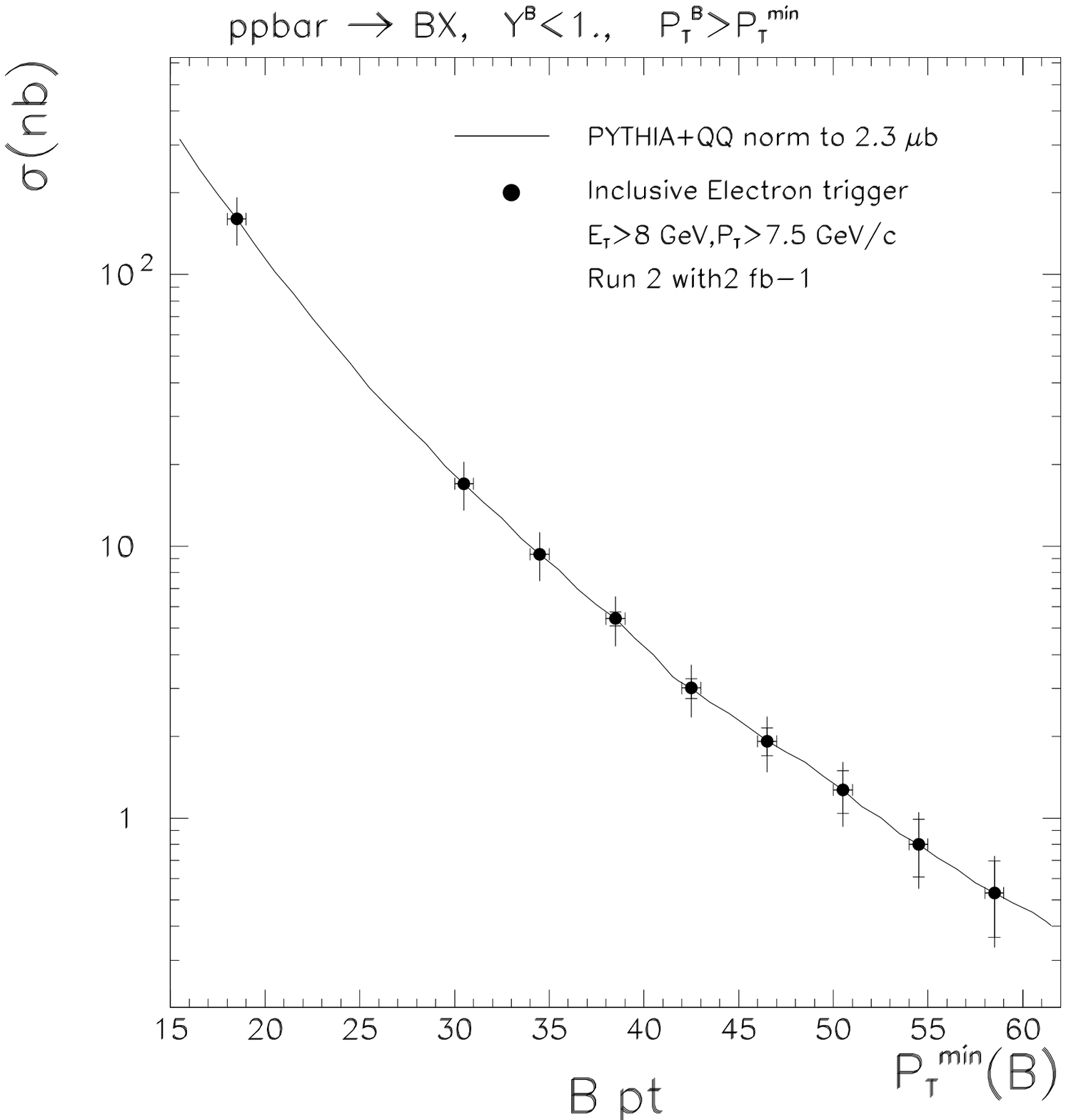}
\put(-395,195){\large\bf (a)}
\put(-165,200){\large\bf (b)}
}
\caption{
(a) 
The Forward-Backward asymmetry $A_{FB}$ in $B^0 \to K^{*0}\mu\mu$ decay
as a function of $s = m_{\mu\mu}^2$ predicted with the Standard Model 
(solid line), SUGRA (dotted) and MIA-SUSY (long-short dashed 
line).
(b)
The predicted $b$~quark production cross section reach as of function of
$B$~meson $p_T$ 
for $B \rightarrow D^0 e X$ assuming 2~fb$^{-1}$ of data.
}
\label{rareprosp}
\end{figure}

Rare $B$ decays provide detailed tests of the flavour structure of the 
Standard Model at the loop level, and as
such provide a complementary probe of new physics to that of direct 
collider searches.
As an example, we discuss the decay $B^0 \to K^{*0}\mu^+\mu^-$. 
The Forward-Backward asymmetry $A_{FB}$ in this decay 
is defined as
\begin{equation}
A_{FB} = 
\frac{N(\cos\Theta>0) - N(\cos\Theta<0)}
{N(\cos\Theta>0) + N(\cos\Theta<0)}
=\frac{N_F-N_B}{N_F+N_B}
\end{equation}
where $\Theta$ is the angle between the direction of the $B^0$ and the
direction of the $\mu^+$ in the rest frame of the $\mu^+\mu^-$ system.
In general $A_{FB}$ depends on the decay
kinematics.  SM calculations predict the distribution of
$A_{FB}$ as a function of the dimuon mass to cross the zero around
$\sqrt{s}=m_{\mu\mu}\sim 2$~\gevcc\ which is stable under various form-factor
parameterizations. Figure~\ref{rareprosp}(a) compares the $A_{FB}$ 
distributions predicted by the Standard Model with several SUSY 
models including SUGRA and MIA-SUSY.
Some new physics models
predict there to be no zero point in the $A_{FB}$ distribution.

CDF expects about 60~$B^0 \to K^{*0}\mu\mu$ event in Run\,IIa which will
allow this decay to be seen at the Standard Model level. However,
precision studies (particularly of the zero point in the 
Forward-Backward asymmetry) will require larger integrated luminosity.  

\newpage
\begin{boldmath}
\subsubsection*{$B$ Production and Spectroscopy}
\end{boldmath}

The discrepancy between the predicted and measured $b$~quark 
production cross section at the Tevatron is one of the largest
disagreements between Standard Model theory and measurement. A
better understand of $b$~quark production in $p\bar p$~collisions,
especially in the high-$p_T$ region as well as $b\bar b$~production
correlations is clearly desirable. CDF plans to
investigate lepton-$D^0$ events to tag $b$-jets in Run\,II.
Figure~\ref{rareprosp}(b) shows
the predicted cross section reach as of function of $B$~meson $p_T$
for $B \ra D^0 e X$ with
a kinematic cut on the electron of $E_T>$ 8~GeV 
and $p_T>$ 7.5~GeV/$c$ assuming 2~fb$^{-1}$ of luminosity. 
The statistical errors are 
predicted from Monte Carlo and scaled by a factor of two 
to include the effect of background.

\vskip 0.4cm
\subsection*{Conclusion}

After a five year upgrade period, the CDF detector is back in operation
taking high quality data in Run\,II 
with all subsystems
functional including a new central tacking 
chamber and silicon vertex detector as well as a hadronic track
trigger and a Time-of-Flight system. The understanding of
the detector is advanced and first physics results have been presented. 
We also
discussed the prospects for $B$~physics in Run\,II, in particular the
measurements of \Bs~flavour oscillations and $CP$~violation in $B$~decays. 
More detailed information about $B$~physics prospects at the Tevatron in
Run\,II can be found in Ref.~\cite{breport}. 
   


\end{document}